\documentclass[prd,twocolumn,superscriptaddress]{revtex4}
\usepackage{graphicx}
\usepackage{amsmath}
\usepackage{amsfonts}
\usepackage{amssymb}
\usepackage{color}
\usepackage{amscd}
\usepackage{bm}
\makeatletter
\renewcommand{\maketag@@@}[1]{\hbox{\m@th\normalsize\normalfont#1}}%
\makeatother
\begin{document}
\title{On the first order corrections to the black hole thermodynamics in higher curvature theories of gravity}
\author{Yong Xiao}
\email{xiaoyong@hbu.edu.cn}
\affiliation{Key Laboratory of High-precision Computation and Application of Quantum Field Theory of Hebei Province,
College of Physical Science and Technology, Hebei University, Baoding 071002, China}

\begin{abstract}
In modified theories of gravity, higher curvature terms may be added to the Einstein-Hilbert action. Conventionally, the effects of the higher curvature terms on the black hole thermodynamics are rather difficult to obtain. In this paper, we show that, at least at the first order level, the corrections to the thermodynamics of the Schwarzschild black hole can be easily obtained, without solving the modified black hole metric. We examine some specific examples, which produces the known results in the literature, as well as those previously unknown. Furthermore, we find some nonperturbative exact results by generalizing the first-order analysis. In particular, a novel relation between the Euclidean integrals of the Einstein--Hilbert term and the higher curvature terms is found and proved.
\end{abstract}
 \maketitle

\section{Introduction}
Einstein gravity is a successful gravitational theory from many aspects. However, one can still consider modifying it at the energy scale $E>100$ MeV, where it has not been fully tested. In fact, in the spirit of effective field theory, it is permissible to add all the possible diffeomorphism invariant curvature terms to the Einstein--Hilbert (EH) action, thus modifying the original Einstein gravity. Indeed, such higher curvature terms can be naturally generated by quantum fluctuations at short distances \cite{Donoghue:1994dn}, or from the low energy limit of the string theory \cite{Boulware:1985wk}. Over the past decades, higher curvature theories of gravity have attracted much attention. In AdS/CFT, these higher curvature terms correspond to the large N corrections to CFTs. And in cosmology, it has a lot of implications on the scenario of early inflation and the present acceleration of the universe \cite{staro,Clifton:2011jh}.

Generally, when the higher curvature terms are added to the EH action, the original Schwarzschild black hole (SBH) can not be the exact solution of the theory anymore. And accordingly, the black hole thermodynamics (i.e., entropy, temperature and radius of the horizon) also receives corrections. For example, in Guass--Bonnet gravity (or more generally the Lanczos-Lovelock gravity), some exact, spherically symmetric black hole solutions have been found, and the black hole thermodynamics has been analyzed \cite{Boulware:1985wk,Cai:2001dz}. Despite the few cases, for most of the higher curvature theories of gravity, it is very difficult to obtain exact black hole solutions. However, one can always solve them perturbatively and iteratively. To be concrete, let us consider the effective action of the form
\begin{align}
I_{eff}= \int d^4 x \sqrt{-g}\left (\frac{R}{16\pi}+\alpha F(g_{\mu\nu},\nabla_\mu,R_{\mu\nu\rho\sigma}) \right),\label{action}
\end{align}
where the first part is the classical EH term, and the second part is the diffeomorphism-invariant higher curvature term, which is constructed from the metric $g_{\mu \nu}$, the operator $\nabla_{\mu}$ and Riemann tensor $R_{\mu\nu\rho\sigma}$. The parameter $\alpha$ controls the overall effects of the higher curvature terms, and its value should be small to avoid a direct conflict with existing experiments.

Conventionally, one should vary the effective action with respect to the metric $g_{\mu \nu}$ in order to get the gravitational field equations. Then the static and spherical symmetric black hole solution is to be solved under the ansatz
\begin{align}
    ds^2=-A(r)dt^2 +\frac{1}{B(r)} dr^2 +r^2 d \Omega, \label{ansatz}
\end{align}
where $A(r)$ is usually written as $A(r)=N(r) B(r)$. As mentioned, the metric can be solved perturbatively around SBH, with the first-order corrections to the original SBH metric denoted by $\alpha\,a(r)$ and $\alpha\,b(r)$. Then the metric at first order is solved and given by
\begin{align}
    A(r)& \doteq 1-\frac{2M}{r}+\alpha\,a(r), \label{mar}\\
    B(r)& \doteq 1-\frac{2M}{r}+\alpha\,b(r). \label{mbr}
\end{align}
 Throughout the paper, we use the symbol ``$\doteq$" to represent that the expression is valid merely at the first order in $\alpha$, omitting the higher order components $\mathcal{O}(\alpha^2)$. Then, with the metric at hand, the effects of higher curvature terms on the black hole thermodynamics can be derived. Concretely speaking, the horizon radius can be obtained from $A(r_h)=B(r_h)=0$; the temperature from the conical singularity of the metric; and the entropy from the Wald formula or other available techniques.

Though the above processes can work in principle, practically it is rather tedious to realize. For most higher curvature terms, the variation problem in deriving the gravitational field equations has already been complicated and even formidable, not to mention solving it. See, for instance, Ref.\cite{Biswas:2013cha,Xiao:2021ewv}. So there is often a long way to go before really getting the corrections to the black hole thermodynamics.

In this paper, we present an interesting fact that, at least at the first order level, the corrections to the SBH thermodynamics can be easily obtained, without solving the modified black hole metric. The paper is organized as follows. First, we describe the procedure of our approach. Then we examine some specific examples, which produces the known results in the literature, as well as those previously unknown. Next, we provide some exact results inspired and obtained by generalizing the first-order analysis. In particular, a novel relation between the Euclidean integrals of the EH term and the higher curvature term is found and proved. We conclude with a summary and discussion, while the explanations of some assertions of the paper are left to the Appendix.

\section{The procedure of our approach}\label{sec2}

We mainly use the Euclidean approach to derive the black hole thermodynamics. In this approach, one should do the Euclidean integral of the action and take it as the partition function of the thermodynamics. The spirit of the section is that the \emph{first-order} partition function can be calculated by using only the \emph{zeroth-order} metric, i.e., the original SBH metric. In the following, we shall describe the procedure, and to avoid diverging too much from the main thread, the proofs of some assertions are left to the Appendix. Below we work in $D=4$ dimensions, but the analysis can be straightforwardly generalized to higher dimensions.

First, we evaluate the Euclidean integral of the EH part of the effective action \eqref{action}, that is,
\begin{align}
    I_1\equiv \beta \int d^3 x \sqrt{-g}\frac{R}{16\pi}. \label{I1def}
\end{align} The temporal integral has given the time periodicity of black hole space-time, which is the inverse of the temperature $\beta\equiv 1/T$. The spatial integral runs from the horizon radius to infinity. For now, we only need to know the existence of the horizon, but don't need to know its precise location. As explained in the Appendix, if $A(r)=B(r)$ in the metric, the integral can be done exactly, which gives
$I_1=\frac{\beta M}{2}+\pi r_h^2-\frac{\beta r_h}{2}$, where $r_h$ is the horizon radius. If $A(r)\neq B(r)$, the formula becomes approximately valid only at the first order in $\alpha$. Thus, whatever the case is, we always have
\begin{align}
   I_1\doteq\frac{\beta M}{2}+\pi r_h^2-\frac{\beta r_h}{2}.\label{eu1}
\end{align}

Second, the Euclidean action implicitly includes the contribution from the Gibbs-Hawking-York (GHY) term or other necessary boundary terms.  For the SBH and its modifications in the higher curvature gravity, we can prove the boundary term always takes the exact form
\begin{align}
    I_2= -\frac{\beta M}{2}.\label{eu2}
\end{align}
Adding $I_1$ and $I_2$ together, there is $I_1+I_2\doteq \pi r_h^2-\frac{\beta r_h}{2}$. In order to reduce the number of variables for the convenience of computation, it can be further re-expressed in the form
\begin{align}
   I_1+I_2\doteq -\frac{\beta ^2}{16 \pi }. \label{act12}
\end{align}
Note that the parameter $\alpha$ doesn't explicitly present here even though it is a first-order expression. In fact, the expression $ -\frac{\beta ^2}{16 \pi }$ is the same partition function for the original SBH in Einstein gravity, and was occasionally borrowed without illustration when Einstein gravity is modified \cite{El-Menoufi:2017kew,Xiao:2021zly}. However, whether and to what extent it could be safely applied were far from clear. In the Appendix, we prove that it is still applicable to the evaluation of $I_1+I_2$ at the first order level, just meeting our current needs. A further analysis can show that it fails at second and higher orders.

Third, we evaluate the contribution from the higher curvature terms, that is,
\begin{align}
    I_3\equiv  \beta \int d^3 x \sqrt{-g}\, \alpha F(g_{\mu\nu},\nabla_\mu,R_{\mu\nu\rho\sigma}). \label{I3def}
\end{align}
Because there has already been a coefficient $\alpha$, we only need to use the original SBH metric to calculate the integrand and do the integration. It yields
\begin{align}
  I_3\doteq\beta \,\alpha F(M,r_h)\doteq\beta \,\alpha F(\beta) \label{I3},
\end{align}
with a little abuse of the notation $F(\cdots)$. In the last step, we used the relation between $M$, $r_h$ and $\beta$ of SBH to reduce the number of variables.

In total, the partition function is
\begin{align}
   I\doteq -\frac{\beta ^2}{16 \pi }+\beta \,\alpha F(\beta) \label{Itotal}.
\end{align}
The energy and entropy of the black hole can be derived by the standard formulas
\begin{align}
    &M(\beta)=-\frac{\partial I}{\partial \beta}, \label{mass}\\
    &S_E(\beta)=\beta M+I. \label{euentr}
\end{align}
The shift of the horizon radius can also be determined. The Wald entropy formula \cite{Wald:1993nt,Iyer:1994ys} provides an independent way to compute the entropy, with the form
\begin{align}
S_{W} =  -2\pi \oint
	 (\frac{\partial \mathcal{L}}{\partial R_{\mu\nu\rho\sigma}})^{(0)} \epsilon_{\mu\nu} \epsilon_{\rho\sigma} d\Sigma. \label{wald}
\end{align}
Taking the derivative of $\mathcal{L}=\frac{R}{16\pi}+\alpha F(g_{\mu\nu},\nabla_\mu,R_{\mu\nu\rho\sigma})$ with respect to Riemann curvature is a standard calculation \cite{Maeda:2010bu}. The EH term always leads to the area entropy $\pi r_h^2$. And as above, when evaluating the output of the higher curvature term, we can straightforwardly employ the curvatures of the original SBH. It gives
\begin{align}
    S_W(r_h) \doteq \pi r_h^2+\alpha F(r_h). \label{waldentr}
\end{align}
The non-trivial agreement between the Euclidean and Wald entropy has been studied, for example, in \cite{Dutta:2006vs}. Thus, by requiring $S_E(\beta)=S_W(r_h)$, the relation between $r_h$ and $\beta$ can be solved. Interestingly, the Wald entropy \eqref{waldentr} looks more like a geometric quantity compared to the thermodynamic Euclidean entropy \eqref{euentr}.

 If needed, the entropy, temperature and horizon radius can also be re-expressed as the functions of $M$, the quantity that can be soundly defined and measured as the conserved energy of the space-time. Obviously, from eq.\eqref{mass}, $\beta$ can be inversely solved as the function of $M$.

 After the above procedures, the information about the first order corrections to the black hole thermodynamics has been revealed.

\section{some examples} \label{sec3}
In this section, we produce results for some concrete examples and compare them with those existing in the literature.

First, consider the example
\begin{align}
    \alpha  F(g_{\mu\nu},\nabla_\mu,R_{\mu\nu\rho\sigma})= \frac{\alpha}{m_\Lambda^2} R^{\mu\nu}_{\ \ \alpha\beta} R^{\alpha\beta}_{\ \ \rho\sigma}R^{\rho\sigma}_{\ \ \mu\nu},
\end{align}
where $m_\Lambda$ denotes the energy scale where the effects of the added term become salient. But we shall set $m_\Lambda=1$ for brevity from now on. According to eqs.\eqref{I3} and \eqref{Itotal}, the partition function can be evaluated as
\begin{align}
I\doteq-\frac{\beta ^2}{16 \pi }+ \alpha  \frac{512 \pi ^4 }{\beta ^2}.
\end{align}
Then, following the procedure of last section, we get the thermodynamic quantities
\begin{align}
  T &\doteq \frac{1}{8 \pi  M}+ \alpha \frac{1}{4 M^5}, \label{curT}\\
  S &\doteq 4 \pi  M^2+\alpha \frac{8 \pi ^2 }{M^2},\label{curS}\\
  r_h &\doteq   2 M- \alpha \frac{10 \pi  }{M^3} \label{curR},
\end{align}
where the second terms in the above expressions represent the first order corrections to the black hole thermodynamics due to the higher curvature term of the action. The results \eqref{curT}, \eqref{curS} and \eqref{curR} are exactly the same as eqs.(16), (18) and (20) of \cite{Calmet:2021lny} which exploited the conventional approach.

For higher and higher curvature terms, the conventional approach becomes much harder to work through. The variation problem in deriving the gravitational field equations has already been a hard work, before really solving the black hole metric and the corresponding thermodynamics. In contrast, our approach is still simple and easily executed. Without too much effort, for the general situation
\begin{align}
    \alpha  F(g_{\mu\nu},\nabla_\mu,R_{\mu\nu\rho\sigma})= \alpha \overbrace{ R^{\mu\nu}_{\ \ \alpha\beta} R^{\alpha\beta}_{\ \ \centerdot\,\centerdot} \cdots R^{\centerdot\,\centerdot}_{\ \ \mu\nu}}^n, \label{rnn}
\end{align}
we find the correction terms as
\begin{align}
T^{(c)} &\doteq \alpha\frac{4^{2-n} (n-2) \left( 2^n+2 (-1)^n \right) M^{1-2 n}}{3 (n-1)} ,\\
S^{(c)} &\doteq \alpha \frac{  2^{9-2 n} \left( 2^n + 2 (-1)^n \right) \pi ^2 M^{4-2 n}}{3 (n-1)},\\
r_h^{(c)} &\doteq \alpha \frac{   2^{5-2 n} [8 (-1)^n- 2^n ( 3 n^2- 3 n-4 ) ] \pi M^{3-2 n}}{3 (n-1)}.
\end{align}
By the way, if using $r_h$ as the variable, instead of $M$, the expression of entropy becomes
\begin{align}
    S\doteq \pi r_h^2+\alpha   2^{n+3} n \pi ^2 r_h^{4-2 n}.
\end{align}
From the above expressions, the case $n=3$ recovers to what we just have given. And the case $n=2$ corresponds to the Gauss-Bonnet gravity (the scalar curvature $R$, Ricci tensor $R_{\mu\nu}$ of SBH vanish and do not contribute at the first order) in $D=4$ dimensions, where the temperature and horizon radius remain unchanged and the entropy correction is a constant.

Next we turn to another example
\begin{align}
     \alpha  F(g_{\mu\nu},\nabla_\mu,R_{\mu\nu\rho\sigma})=\alpha \sum_{n=0}^{\infty} f_n R^{\mu\nu\rho\sigma}\Box^n R^{\mu\nu\rho\sigma}, \label{boxes}
\end{align}
where the right hand side comes from a Taylor expansion of some operator $f(\Box)=\sum f_n \Box^n$ \cite{Biswas:2013cha}. The gravitational field equations are extremely involved and complicated, and its first-order corrections to the black hole thermodynamics were studied recently in \cite{Xiao:2021ewv} using the conventional approach. Once again, now we can avoid the arduous computations therein. Following eqs.\eqref{I3} and \eqref{Itotal}, the partition function can be evaluated as
\begin{align}
    I \doteq -\frac{\beta ^2}{16 \pi }+64 \pi ^2 \alpha f_0   - \alpha f_1\frac{1536 \pi ^4 }{\beta ^2}+\alpha f_2 \frac{49152 \pi ^6 }{\beta ^4}+\cdots.
\end{align}
Thus the thermodynamic quantities are
 \begin{align}
T &\doteq \frac{1}{8 \pi M } - \frac{3 \alpha f_{1}}{4 M^5} + \frac{3 \alpha f_{2}}{4 M^7}  +\cdots,\label{temmis}\\
S &\doteq 4 \pi  M^2 +64\pi^2 \alpha f_{0} - \frac {24 \pi^2 \alpha f_{1}}{M^2} + \frac {12 \pi^2 \alpha f_{2}}{M^4}\cdots,\\
r_h &\doteq  2M + \frac{6 \pi \alpha  f_{1} }{M^3} - \frac{15 \pi \alpha f_{2} }{M^5}  +\cdots.
\end{align}
The general expressions for the thermodynamic quantities up to any value of $n$ of eq.\eqref{boxes} haven't been found, but the interested readers can rapidly generate further results by a simple program following our procedure.

\section{Some exact results}

\subsection{An exact relation between $I_1$ and $I_3$}
From our approach, a lot of data about the thermodynamic quantities for various higher curvature theories can be easily generated and accumulated. This gives us a chance of finding some rules or patterns, which may still be applicable to higher orders in $\alpha$. In particular, we find an interesting relation between $I_1$ and $I_3$. And it is an exact relation, meaning that it is applicable to the exact black hole solutions, or perturbative solutions to any order.

As an example, substituting eqs.\eqref{curT} and \eqref{curR} into the expression for $I_1$ and $I_3$, we find
\begin{align}
    I_1 &\doteq -\frac{\beta ^2}{16 \pi }-(-\frac{\beta M}{2}) \doteq \alpha \frac{8 \pi ^2 }{M^2} \\
    I_3 &\doteq \alpha  \frac{512 \pi ^4 }{\beta ^2} \doteq \alpha \frac{8 \pi ^2 }{M^2}.
\end{align}
Obviously, there is $I_1=I_3$. At first sight, it looks more like a coincidence, since the definitions for $I_1$ and $I_3$ look so different. However, it is not that curious from another perspective. For the original SBH in Einstein gravity, $I_1=0$; after the higher curvature term is added, the black hole metric gets shifted a little, causing $I_1$ nonzero and closely related to the value of $I_3$.

Generally, we find
\begin{align}
 I_1=(m-2)I_3,
 \end{align} where $m$ counts the number of curvatures $R$, $R_{\mu\nu}$ $R_{\mu\nu\rho\sigma}$, and also half the number of the operator $\nabla_\mu$'s that are explicitly included in $ F(g_{\mu\nu},\nabla_\mu,R_{\mu\nu\rho\sigma})$. Taking a random example, $R R_{\alpha \beta }R^{\alpha \nu\rho\sigma} \Box  R^{\beta}_{\ \nu\rho\sigma}$, one counts $m=5$. We further explore this issue for SBHs in $D \geq 4$ dimensions, and the formula turns out to be
\begin{align}
I_1=-\frac{D-2m}{D-2}I_3. \label{I13}
\end{align}
A careful examination of various examples shows that the formula \eqref{I13} is actually an exact formula applicable to any orders, so there must be some mathematical structure hidden behind it. Surely, we can figure out a proof for the formula \eqref{I13} as below.

Varying the action \eqref{action}, we find that the gravitational field equations always have the structure
\begin{align}
\frac{1}{16 \pi} (R^{\mu \nu}-\frac{1}{2} g^{\mu \nu}R)=-\alpha ( f^{\mu \nu}-\frac{1}{2} g^{\mu \nu}F) - \alpha \nabla^\mu K^\nu, \label{eofpr}
\end{align}
where $F(g_{\mu\nu},\nabla_\mu,R_{\mu\nu\rho\sigma})$ is abbreviated as $F$. On the right hand side, $\frac{1}{2} g^{\mu \nu}F$ comes from  $\delta(\sqrt{-g})\, F$, and $-\alpha ( f^{\mu \nu}+ \nabla^\mu K^\nu)$ comes from $\sqrt{-g}\,\delta F$. In addition, $f_{\mu\nu}$ has a useful property $g_{\mu \nu}f^{\mu\nu}=m F$, and $\nabla^\mu K^\nu$ represents a total derivative term.

In order to understand eq.\eqref{eofpr} better, consider the typical example $F=R R_{\alpha \beta }R^{\alpha \nu\rho\sigma} \Box  R^{\beta}_{\ \nu\rho\sigma}$, and write it as
\begin{align}
\begin{split}
    F=& g^{\mu_1 \mu_2} g^{\nu \mu_3}g^{\rho \mu_4}g^{\sigma \mu_5}  g^{\mu_6 \mu_7}\\
    \bullet\,  & R_{\mu_1 \mu_2} R_{\alpha \beta} R^{\alpha}_{\ \mu_3\mu_4\mu_5} \nabla_{\mu_6} \nabla_{\mu_7} R^{\beta}_{\ \nu\rho\sigma}.
    \end{split}
\end{align}
Then $\delta F  $ can be written in the form $- ( f^{\mu \nu}+ \nabla^\mu K^\nu) \delta g_{\mu \nu}$. The five terms deduced from the five different $\delta g^{\alpha \beta}= -g^{\alpha \mu} g^{\beta\nu} \delta g_{\mu\nu}$ are collected in $- f^{\mu \nu}$. The terms deduced from $\delta R^{\alpha \beta }$, $\delta R^{\beta}_{\ \nu\rho\sigma}$ and $\delta \nabla_\mu\sim \delta \Gamma$ are total derivative terms and collected in $-\nabla^\mu K^\nu $. Moreover, when contracted with $g_{\mu \nu}$, each of the five terms in $f^{\mu \nu}$ gives back to the same form $F$, so $g_{\mu \nu}f^{\mu\nu}=5 F$ in this case.

Let's continue to do the proof. With eq.\eqref{eofpr} at hand, contracting it with $g_{\mu \nu}$ and do the Euclidean integral from the horizon radius to infinity, we get
\begin{align}
  (1-\frac{D}{2})  \int_\mathcal{M} \frac{ R}{16 \pi}=- (m-\frac{D}{2}) \int_\mathcal{M} \alpha F- \int_\mathcal{M} \alpha \nabla_\mu K^\mu,\label{pr1}
   \end{align}
where  $\int_\mathcal{M} \centerdot \equiv \beta \int d^{D-1} x\sqrt{-g}\,\centerdot$. In fact, we can get rid of the last term, because it simply vanishes \footnote{In the case $m \neq \frac{D}{2}$, $ \int_\mathcal{M} \nabla_\mu K^\mu$ is dimensionful by direct counting, so it must vanish. If not, it would be a function of $M$; imagining adding $ \alpha \int d^4 x \sqrt{-g} \nabla_\mu K^\mu$ to the action, it would not change the black hole solutions while changing the thermodynamics, which is physically unreasonable. The case $m=\frac{D}{2}$ is another story. Now $ \int_\mathcal{M} \nabla_\mu K^\mu$ is dimensionless, so in principle it could be some number denoted by $c_1$. Because of eq.\eqref{pr1}, $I_1= \frac{2 \alpha}{D-2}c_1$. Then the partition function is $I\doteq -\frac{\beta M(\beta)}{D-2}+\text{const.}$, from which we can solve the thermodynamics and finally get $I_1= 0$. So there must be $c_1=0$. The fact of $I_1\equiv \int_M \frac{R}{16\pi}=0$ ($m=\frac{D}{2}$) is trivial in $D=4$ dimensions, because $\mathcal{O}(R^2)$ terms don't
t change the SBH metric where $R=0$. And this is non-trivial in higher dimensions, since the SBH metric can be changed along with $R\neq 0$.}. Thus, eq.\eqref{pr1} can be re-expressed as
\begin{align}
   \int_\mathcal{M} \frac{ R}{16 \pi}=-\frac{D-2m}{D-2} \int_\mathcal{M}  \alpha  F(g_{\mu\nu},\nabla_\mu,R_{\mu\nu\rho\sigma}),
\end{align}
which finishes our proof of eq.\eqref{I13}, applying the definition of $I_1$ and $I_3$.

Clearly, when $F(g_{\mu\nu},\nabla_\mu,R_{\mu\nu\rho\sigma})$ is a sum of individual higher curvature terms $\mathcal{O}(R^m)$, the formula becomes
\begin{align}
    I_1=-\sum_m\frac{D-2m}{D-2}I_3^{(m)} .
\end{align}

\subsection{On the Gauss-Bonnet gravity in higher dimensions}

The Gauss-Bonnet gravity is simple in that its field equations only contain second derivatives of the metric. Thus, the exact black hole solutions are not hard to obtain in higher dimensions \cite{Boulware:1985wk,Cai:2001dz}, and there is no necessity to derive the thermodynamics from our approach. But maybe it is still interesting to see how the exact thermodynamic quantities emerge with only partial information of the metric. We shall work in $D=5$ dimensions as the example.

Assume we have known $A(r)=B(r)$ and their zeroth order approximation, i.e., the original SBH metric $A(r)=B(r)
\simeq 1-\frac{8 M}{3 \pi  r^2}$. Note that the $D=5$ generalizations of eqs.\eqref{eu1} and \eqref{eu2} are
\begin{align}
    I_1 = \frac{\beta M}{3}+\frac{\pi ^2 r_h^3}{2}-\frac{3}{8} \pi  \beta  r_h^2, \label{gbI1}
\end{align}
and $I_2=-\frac{\beta M}{3}$. The Gauss--Bonnet term is
\begin{align}
\alpha F(g_{\mu\nu},\nabla_\mu,R_{\mu\nu\rho\sigma})=\frac{\alpha}{16 \pi}(R^2-4 R_{\mu \nu}R^{\mu \nu}+R_{\mu \nu \rho \sigma }R^{\mu \nu  \rho \sigma }).
\end{align}
Then the Euclidean integral gives
$\beta\frac{3}{4} \pi  \alpha  [2 r (A(r)-1) A'(r)+A(r)^2-2 A(r)]^{\infty}_{r_h}$.
As $r\rightarrow \infty$, the expression reduces to $-\beta \frac{3\pi \alpha}{4}$. At horizon, it can be simplified by $A(r_h)=0$, $A'(r_h)=4\pi/\beta$. Then the subtraction between the infinity and the horizon radius leads to $I_3=6 \pi ^2 \alpha  r_h-\beta \frac{3\pi \alpha}{4}$. In total, the partition function is
\begin{align}
I_1+I_2+I_3=\frac{\pi ^2 r_h^3}{2}+6 \pi ^2 \alpha  r_h - \beta  \left( \frac{3 \pi  \alpha }{4}+\frac{3 \pi  r_h^2}{8}\right).
\end{align}
 Comparing with the standard form $\ln Z=S-\beta M$, we read
 \begin{align}
    M= \frac{3 \pi  r_h^2}{8}+\frac{3 \pi  \alpha }{4},\label{gbM}\\
    S=\frac{\pi ^2 r_h^3}{2}+6 \pi ^2 \alpha  r_h.
 \end{align}
The temperature is calculated by
 \begin{align}
    T= \frac{\partial M}{\partial S}=\frac{r_h}{8 \pi  \alpha +2 \pi  r_h^2}.\label{gbT}
 \end{align}
  These are surely the exact results known for the Gauss-Bonnet black hole \cite{Cai:2001dz}.

Incidentally, we verify the formula \eqref{I13} again. Substituting  eqs.\eqref{gbM} and \eqref{gbT} into the expressions of $I_1$ and $I_3$, and using $r_h$ as the independent variable, we have $I_1=\frac{2 \pi ^2 \alpha ^2}{r_h}-\frac{3}{2} \pi ^2 \alpha  r_h $ and $I_3=\frac{9}{2} \pi ^2 \alpha  r_h-\frac{6 \pi ^2 \alpha ^2}{r_h} $. So there is
\begin{align}
I_1=-\frac{1}{3}I_3,
\end{align}
which exactly matches the case $D=5$, $m=2$ of eq.\eqref{I13}.

\section{conclusions}
In this paper, we present an interesting fact that, at least at the first order level, the corrections to the thermodynamic quantities of SBH, due to higher curvature terms in the action, can be easily obtained. We have recovered some known first-order results existing in the literature, and produced more results which may be hard (if not possible) to obtained from the conventional approach. Actually, in previous literature, the attention was mainly focused on the effects of the curvature-squared and cubic-curvature terms, while those for much higher curvature terms were seldom achieved. In contrast, now we have provided a convenient and rapid approach to get the corrections to the black hole thermodynamics for an arbitrarily given diffeomorphism-invariant higher curvature term, for instance, see eq.\eqref{rnn}.

By hindsight, the reason why our approach works well is that the Schwarzschild solution extremizes the Einstein-Hilbert action. The Euclidean action $I_1+I_2$ of the original Schwarzschild solution in canonical ensemble is $-\frac{\beta \bar{M}(\beta)}{2}$, where $\bar{M}(\beta)=\frac{\beta}{8\pi}$. Because it is an extremum of the thermodynamics, the form will not change at the first order, even the metric has already been perturbed somehow. This explains the success of our usage of eq.\eqref{act12}, that is $I_1+I_2\doteq -\frac{\beta ^2}{16 \pi }$. The corrections to the expression only emerge from the second order of the perturbations.

According to the same spirit, this approach can be generalized to other types of black holes. For the rotating black holes, the corrections to the Kerr black hole thermodynamics due to higher curvature terms of $\mathcal{O}(R^3)$ and $\mathcal{O}(R^4)$ have been analyzed in \cite{Reall:2019sah}. The Euclidean action $I_1+I_2$ can be approximated as $-\frac{\beta \bar{M}(\beta, \Omega)}{2}$, where $\bar{M}$ is the mass of the unperturbed Kerr black hole as the functions of the given $\beta$ (inverse temperature) and $\Omega$ (angular velocity). See \cite{Reall:2019sah} for details.

As for the AdS black holes, it was noticed in \cite{Caldarelli:1999ar,Landsteiner:1999gb} that the corrections to the thermodynamics can be obtained without knowing the perturbed metric. But it was only a direct observation without a clear explanation there. From our perspective, their success is because the reference AdS background has not been modified. For general higher curvature terms, the cosmological constant $\Lambda$ may get ``renormalized" to be some $\Lambda_{eff}$. Indeed, if this effect has been considered and compensated (to calibrate the subtracted background), the approach still nicely works \footnote{For hyperbolic black holes, the extremal black hole is used as the reference background \cite{Caldarelli:1999ar}. However, we find that the effectiveness of the approach depends on the behaviors of the background at infinity. So here it is still the AdS background that matters.}.

Besides, we have found a nonperturbative exact relation \eqref{I13} between the Euclidean integrals of the EH term $I_1$ and higher curvature term $I_3$. This is an unexpected finding from the accumulated data, but we also have proved it. At first sight, it is surprising that the two quantities have such a simple relation, especially noting that it holds for general higher curvature terms.

\ \
\section*{Acknowledgments}
YX would like to thank Harvey S. Reall and Jorge E. Santos for extensive discussions about the corrections to the Kerr black hole thermodynamics, and also Longting Zhang for helpful discussions. This work was supported in part by NSF of Hebei province with Grant No. A2021201022.

\section*{Appendix}

In this Appendix, we provide a brief proof for eqs.\eqref{eu1}, \eqref{eu2} and \eqref{act12}.

 In the case $A(r)=B(r)$, $I_1\equiv \beta  \int d^3 x \sqrt{-g} \frac{R}{16 \pi}$ can be integrated out explicitly. The outcome is $-\frac{1}{4} \beta  [r \left(r A'(r)+2 A(r)-2\right)]^{\infty}_{r_h}$, where $A(r)=1-\frac{2M}{r} +\cdots$, with ``$\cdots$" representing the $r^{-n}$ ($n\geq 2$) corrections to the SBH metric due to the higher curvature terms. As $r\rightarrow \infty$, $r\left(2A(r)-2 \right)\rightarrow -4M$, $r^2 A'(r)\rightarrow 2M$, so the expression gives $\frac{\beta M}{2}$. At the horizon, it can be simplified by $A(r_h)=0$,
$A'(r_h)=4 \pi/{\beta}$. Then, the subtraction between the infinity and the horizon radius leads to $I_1=\frac{\beta M}{2}+\pi r_h^2-\frac{\beta r_h}{2}$. In the case $A(r)\neq B(r)$, unfortunately the integration can't yield exact expressions, so the analysis can only be done at the first order in $\alpha$. Consequently, in general, the above formula of $I_1$ holds only at the first order level. Because the steps are more involved, we don't intend to present it here, but leave it as an exercise.

Next we explain that the boundary terms lead to $I_2= -\frac{\beta M}{2}$. The boundary terms are important to ensure the field equations are well-defined. In particular, the boundary term corresponding to the EH term is the GHY term. And the boundary terms corresponding to the higher curvature terms are often complicated and not generally known. So, in order to get the contributions from the boundary terms, we adopt another strategy. We first consider the black holes in the AdS space-time. There is a subtraction procedure for calculating the Euclidean action of AdS black holes without the necessity of considering the boundary terms, even if higher curvature terms have been included in the action \cite{hawkingpage,Dutta:2006vs}. Then, by taking the limit $\Lambda \rightarrow 0$, we will obtain the Euclidean action for the asymptotic-flat black holes and identify the contributions of the boundary terms. Concretely speaking, for Einstein gravity, the boundary term can be calculated by $ \beta \lim\limits_{r\rightarrow\infty}
[\frac{  r^3\Lambda}{6}- (1+\frac{3 M}{  r^3  \Lambda} ) \frac{ r^3 \Lambda}{6}  ] =-\frac{\beta M}{2}$. When higher curvature terms are added, the result can be sketched as the form  $-\beta  (\frac{M}{2} + \alpha \Lambda M +\cdots )$. Taking the limit  $\Lambda\rightarrow 0$,  we get neatly  $I_2= -\frac{\beta M}{2}$.

In the end, we have $I_1+I_2\doteq \pi r_h^2-\frac{\beta r_h}{2}$. Assume $r_h \doteq 4\pi\beta +\alpha f_1(\beta)$ and substitute it into the expression to reduce the number of variables. Then the unknown function $f_1(\beta)$ cancels; the formula becomes $I_1+I_2\doteq -\frac{\beta ^2}{16 \pi }$.

\end{document}